\documentclass[10pt,conference]{IEEEtran} 

\IEEEoverridecommandlockouts

\usepackage{cite}
\usepackage{amsmath,amssymb,amsfonts}
\usepackage{graphicx}
\usepackage{textcomp}
\usepackage{xcolor}
\usepackage{siunitx}
\usepackage{booktabs}
\usepackage{float}
\usepackage{dblfloatfix}
\usepackage{epsfig}
\usepackage{caption}
\usepackage{url}
\usepackage{algorithm}
\usepackage[noend]{algpseudocode}

\usepackage{todonotes}

\usepackage{soul}

\usepackage[utf8]{inputenc}
\usepackage{booktabs,caption}
\usepackage[flushleft]{threeparttable}
\usepackage{multirow}
\usepackage{dblfloatfix}
\usepackage{comment}

\def\BibTeX{{\rm B\kern-.05em{\sc i\kern-.025em b}\kern-.08em
    T\kern-.1667em\lower.7ex\hbox{E}\kern-.125emX}}

\begin{document}
\title{Sus-STELLAR: A Sustainable Synthesizable Solution to EM/Power SCA and Malicious Voltage Drop-based Attack}
\title{\color{black}{R-STELLAR: A Resilient Synthesizable SCA Protection with built-in Attack-on-Countermeasure Detection}}

\title{\color{black}{R-STELLAR: A Resilient Synthesizable Signature Attenuation SCA Protection on AES-256 with built-in Attack-on-Countermeasure Detection}}

\author{\IEEEauthorblockA{Archisman Ghosh\IEEEauthorrefmark{1},
Dong-Hyun Seo\IEEEauthorrefmark{1},
Debayan Das\IEEEauthorrefmark{2},
Santosh Ghosh\IEEEauthorrefmark{3},
and~Shreyas~Sen\IEEEauthorrefmark{1}} 
\IEEEauthorblockA{\IEEEauthorrefmark{1}School of Electrical and Computer Engineering,
Purdue University, West Lafayette, IN, USA} 
\IEEEauthorblockA{\IEEEauthorrefmark{2}Indian Institute of Science, Bangalore, India}
\IEEEauthorblockA{\IEEEauthorrefmark{3}Intel Labs, Intel Corporation, Hillsboro, OR, USA}
}
\maketitle
\begin{abstract}
Side-channel attacks (SCAs) remain a significant threat to the security of cryptographic systems in modern embedded devices. Even mathematically secure cryptographic algorithms, when implemented in hardware, inadvertently leak information through physical side-channel signatures such as power consumption, electromagnetic (EM) radiation, light emissions, and acoustic emanations. Exploiting these side channels significantly reduces the attacker’s search space. 
In recent years, physical countermeasures have significantly increased the minimum traces-to-disclosure (MTD) to 1 billion. {\color{black}Among them, signature attenuation is the first method to achieve this mark. Signature attenuation often relies on analog techniques, and digital signature attenuation reduces MTD to 20 million, requiring additional methods for high resilience. We focus on improving the digital signature attenuation by an order of magnitude (MTD 200M).} 
Additionally, we {\color{black} explore possible attacks against signature attenuation countermeasure. We introduce a Voltage-drop Linear-region Biasing (VLB) attack technique }that reduces the MTD to over 2000 times less than the previous threshold. This is the first known attack against a physical side-channel attack (SCA) countermeasure. We have implemented an attack detector with a response time of 0.8 milliseconds to detect such attacks, {\color{black} limiting SCA leakage window to sub-ms, which is insufficient for a successful attack}.
\end{abstract}

 \begin{figure}[!ht]
  \centering
   \includegraphics[width=0.45\textwidth]{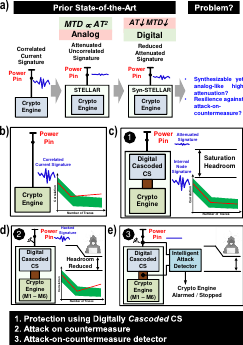}
   \caption{{a) Prior state-of-the-art using signature attenuation techniques. b) Unprotected AES can be attacked using power SCA. c) This work protects against power SCA using digitally {\bf{cascoded}}
   current source. d) {\color{black} A Voltage drop-based Linear-region Biasing attack is explored using a signature attenuation countermeasure. e) The implemented attack detectors can detect this attack for the resilience of signature attenuation countermeasure. Key contributions are tabulated below.}}}
   \label{motivation_1}
 \end{figure}
\begin{figure*}[!ht]
  \centering
   \includegraphics[width=0.98\textwidth]{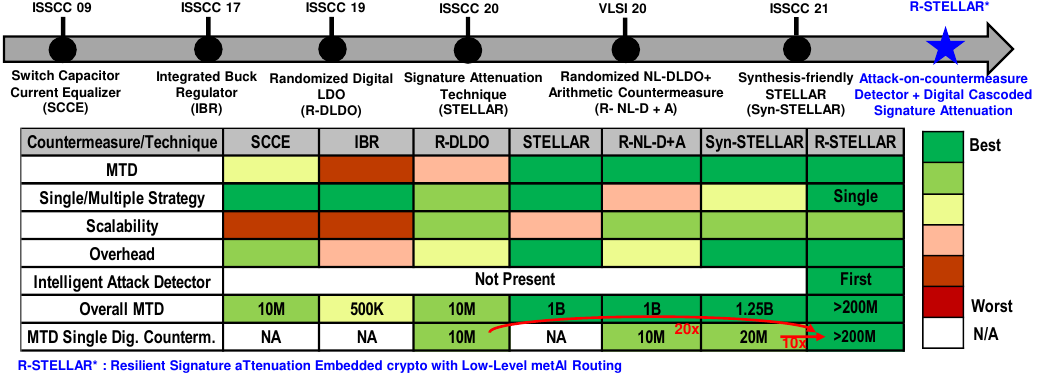}
   \caption{{State-of-the-art circuit level countermeasures. This work brings the benefit of cascoded current sources in the digital domain for high security, even being scalable. 
   }}
   \label{motivation_2}
 \end{figure*}
\section{Introduction}\label{sec:intro}
{\color{black} Cryptographic algorithms are designed to be secure based on mathematical principles. However, they can unintentionally reveal sensitive side-channel information. This type of leakage typically happens due to power correlations, electromagnetic emissions, timing, and variations in cache accesses. Such leaks are a serious security risk for integrated circuits.}
Even the attack complexity of the once-thought-highly secure AES-256 is now reduced to $2^{13}$ from $2^{256}$ due to side-channel analysis (SCA), making it highly vulnerable. Recent observations reveal that the AES-256 key can be intercepted even from a distance using a low-cost EM probe without detailed knowledge of the circuit or PCB implementation.
An adversary monitors the information to exploit this leakage and correlates it against a statistical model constructed using secret key guesses. Correlation attacks utilize Hamming weight (HW) or Hamming distance (HD) models to estimate the switching activities of internal nodes within a cryptographic engine. Depending on the strength of the underlying power model and the availability of power signatures, a correct key guess yields correlation peaks, revealing portions of the secret key.
An alternative analysis for evaluating side-channel vulnerabilities in crypto hardware is the test vector leakage assessment (TVLA) [18]. This analysis estimates model-independent information leakage by applying Welch’s $|t|$-test to a set of fixed and random plaintext vectors. If the resulting $|t|$-score exceeds a heuristic threshold of 4.5, the device is considered to exhibit meaningful leakage. 

The research community has {\color{black}been exploring} various countermeasures in response to the emergence of side-channel attacks (SCA). 
Architectural countermeasures involve heterogeneous S-boxes, arithmetic masking, and multiplicative masking. These methods aim to enhance security by introducing complexity and obfuscation at the architectural level.
In contrast, generic and physical countermeasures address vulnerabilities at the physical implementation level. Examples include randomized series low-dropout (LDO) regulators as well as analog and digital signature attenuation circuits. Some countermeasures combine multiple approaches to achieve robust protection against side-channel leakage. Our work focuses on a high-attenuation technique based on a digitally \textit{cascoded} current source, leveraging a single generic approach.
{\color{black} Security remains a dynamic challenge akin to a cat-and-mouse game. Attackers have questioned the efficacy of certain architectural countermeasures under specific circumstances.} However, attacks against physical countermeasures remain unexplored. We successfully {\color{black} explored an attack on} a physical countermeasure for the first time.
Contributions of this work are three-fold, as shown in Fig.~\ref{motivation_1}:
\begin{itemize}
\item We introduce a signature attenuation technique using a digital cascoded current source, namely Resilient Signature aTtenuation Embedded crypto with Low-Level metAl Routing (R-STELLAR). This technique achieves $20\times$ improvement in minimum-traces-to-disclosure (MTD) compared to single generic, digital, and physical countermeasures.
\item Additionally, we explore an attack{\color{black}, namely Voltage-drop Linear-region Biasing Attack (VLB)} on physical countermeasures, significantly reducing MTD from 200 million traces to 105K traces.
\item Finally, we propose {\color{black} an attack detection mechanism} crucial for the practical adoption of physical countermeasures in industry.
\end{itemize}
The subsequent sections of this paper are structured as follows: In Section II, we delve into related research on countermeasures against power and electromagnetic (EM) side-channel attacks, along with potential vulnerabilities. Section III provides an in-depth analysis of the circuit architecture. In Section IV, we outline the proposed attack strategy and its corresponding mitigation technique. Section V presents the measurement setup, results, and integrated circuit (IC) specifications. Finally, we conclude the paper in Section VI.
\section{Related Works}\label{sec:related_works}
This study enhances the existing state-of-the-art in single digital-friendly countermeasures by a factor of 20, achieved through utilizing a cascoded current source. Additionally, we investigate VLB attack and its corresponding detection mechanism within the same countermeasures. Before delving into the details, we will provide a brief overview of the existing literature.
\subsection{Power/EM SCA countermeasure}
Power \& EM SCA countermeasures can be classified into three categories: a) architectural countermeasures, b) logic-level countermeasures, and c) physical or circuit-level countermeasures. Architectural countermeasures include algorithmic shuffling \cite{yu_aes_2012}, multiplicative masking \cite{golic2002multiplicative}. Algorithmic shuffling rearranges cryptographic operations to disrupt the correlation between power consumption and sensitive data; however, it has limited capability against side-channel attacks as limited operations are shuffled. Time-domain SNR is still high enough to be correlated. Logic-level countermeasures mostly compensate power to gain resilience against power side-channel. WDDL\cite{hwang_aes-based_2006}, SABL\cite{bucci_three-phase_2006}, Dual Rail Precharge Logic\cite{tiri_dynamic_2002}, boolean masking\cite{poschmann_side-channel_2011} are example of the logic-level countermeasures. These solutions are mostly synthesizable; however, they suffer from high power \& area overhead ($>2\times$); hence, they may not be preferred within the scope of area and energy-constrained secure IoT devices. {
\color{black} Recently explored circuit-level countermeasure promises lower overhead while being generic.}
\subsection{\color{black} Circuit-level Countermeasures}
Circuit-level countermeasures ~\cite{ghosh202136,das_et_al_273_2020,ghosh2021syn,he_253_2020,ghosh2022digital,ghosh2023power} solves {\color{black}the problem of practicality} as overhead is significantly less than architectural or logic-level countermeasures. This leads to a recent thrust of circuit-level/physical countermeasures against SCA. The progression of the physical countermeasures is shown in Fig.~\ref{motivation_2}. One popular state-of-the-art countermeasure is switch capacitor current equalizer (SCCE) \cite{tokunaga_secure_2009, tokunaga_securing_2010}. SCCE reaches $>10M MTD$ by supplying the AES with three parallel capacitors and bypassing the information-sensitive leakage to a DC bias. However, this solution suffers from 2$\times$ performance overhead due to large droop caused in the capacitors. Voltage regulator-based solutions include Integrated Buck Regulator (IBR\cite{kar_8.1_2017, kar_reducing_2018})-based solution and series LDO with Loop Randomization (R-DLDO\cite{singh_enhanced_2020}). {\color{black}They provide medium security ($<10M MTD$) due to obfuscation created by different randomization techniques.} However, IBR has large passives (note that MiM cap often radiates meaningful information in terms of EM emanation). Digital LDO inherently leaks critical information as voltage compensation follows the instantaneous current drawn by the crypto-engine. Digital LDO with noise injection and voltage/frequency modulation reaches 6.8M MTD against SCA, although LDO is a high-overhead solution for SCA. Cascade of NL-LDO with arithmetic countermeasures achieves ($>1B\ MTD$) high security against CPA. However, it suffers from high overhead due to LDO and is not generic due to arithmetic countermeasures \cite{kumar2020sca, kumar2021time}. 
\subsection{\color{black}Signature Attenuation Countermeasures}
Signature aTtenuation Embedded crypto with Low-Level metAl Routing (STELLAR) \cite{das_et_al_273_2020} achieves high MTD by using an analog cascoded current source as a power delivery circuit, which provides high attenuation due to its high output impedance. This solution achieves $>1B\ MTD$ for the first time but is not synthesis-friendly. Syn-STELLAR \cite{ghosh2021syn} proposes a scalable signature attenuation-based solution that provides similar MTD ($>1.25B\ MTD$) by cascading two solutions, namely Digital Signature Attenuation Circuit (DSAC) and Time-Varying Transfer Function (TVTF). DSAC does not provide high attenuation compared to CDSA as the synthesizable realization of CS replicates source degenerated structure instead of cascaded structure, contributing to lower attenuation. Additional RO randomization along with TVTF helped to achieve similar security (w.r.t CDSA) at the cost of high overhead. Our solution (namely R-STELLAR: Resilient Signature aTtenuation Embedded crypto with Low-Level metAl Routing) brings the benefit of analog cascoded signature attenuation in the digital domain to achieve high attenuation, hence high MTD ($>200M\ MTD$) against SCA. {\color{black} These solutions use lower metal layer routing to reduce EM leakage.}   
 \begin{figure*}[!t]
  \centering
   \includegraphics[width=0.98\textwidth]{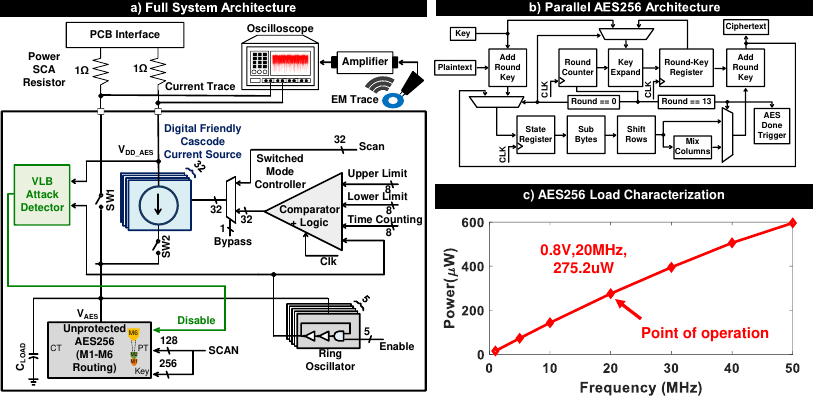}
   \caption{{a) The full system architecture of R-STELLAR. b) Parallel AES-256 archirecture. c) Load characterization of AES.}}
   \label{full_system_archi}
 \end{figure*}
\subsection{Attack against Countermeasure}
Security is {always a strategic contest} between attackers and cryptographers. Advancements in one countermeasure may open another avenue for attack. Historically, countermeasures for square-and-multiply algorithms of RSA scheme against simple power analysis (SPA) have been attacked using Differential Power Analysis three decades ago~\cite{kocher_differential_1999}. Another instance is when exponent randomization-based countermeasures of RSA~\cite{coron1999resistance} have been attacked~\cite{fouque2008carry}. Masking is a provably secure technique. However, different masking techniques of AES have been exploited using higher order attacks or Fault Injection attacks~\cite{saha2023non}. These attack-defense-attack-based explorations of different countermeasure strategies are often explored in the standard crypto community. The recent gamut of physical countermeasures should be tested well against different types of attack strategies. As these countermeasures frequently come from circuit knowledge, {attackers with knowledge of the circuit can increase the probability of attack}. Hence, it is impossible to popularize generic and circuit-level countermeasures without detailed stress testing. Until now, no approach exists to evaluate the countermeasures against new attacks. For the first time, we have explored an attack possibility on physical countermeasures and suggest an attack detector circuit that can detect such an attack through experimental evaluation. This type of attack detector is necessary to sustain the generic countermeasures. We believe this approach will {\color{black}help us increase trust and applicability in physical countermeasures}. Notably, This attack is a demonstration of a signature attenuation-based circuit but can be extended to different physical countermeasures as well, which can be explored as part of future works.    
  \begin{figure*}[!ht]
  \centering
   \includegraphics[width=0.98\textwidth]{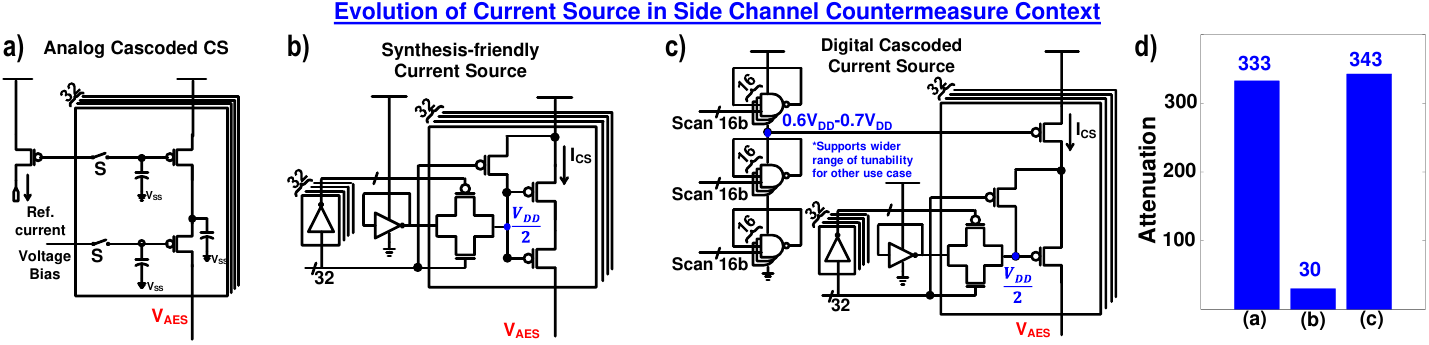}
   \caption{{Progression of signature attenuation circuit from analog to digital domain: a) analog cascoded CS b) digital source degenerated CS which is scalable but provides low attenuation c) digital cascoded current source providing very high attenuation in digital domain d) attenuation by using architecture (a)-(c).}}
   \label{progression_of_CS}
 \end{figure*}
\section{\color{black} R-STELLAR Countermeasure Design} \label{tehnique_against_SCA}
Fig. ~\ref{full_system_archi}(a) presents the full system architecture. The full system architecture consists of a digital cascoded current source (DCCS), multiple scan-controlled parallel ring oscillators (RO) as the bleed path similar to ~\cite{ghosh2021syn}. The bleed path bypasses the delta changes in the supply current, thereby stabilizing the $V_{AES}$ node voltage by providing local negative feedback and hiding small key-dependent current changes. Simultaneously, the RO-bleed is the input of the global feedback (switch mode controller), which is a slow loop that compensates for PVT variation or sudden changes in the crypto current due to frequency variation of the encryption engine. We will discuss DCCS and the Global feedback loop in detail in the following subsections. {\color{black} Parallel AES-256 is used as an example crypto engine as shown in Fig.~\ref{full_system_archi}(b). Load characteristics is shown in Fig.~\ref{full_system_archi}(c). We will discuss AES architecture and load characteristics briefly in section V for continuity. }
\subsection{Digital Cascoded Current Source (DCCS)}
The Digital Cascoded Current Source (DCCS) is crucial in mitigating power and electromagnetic (EM) side-channel attacks. In previous work, Das et al. ~\cite{das_et_al_273_2020} employed an analog cascoded current source, achieving high attenuation (and enhanced security against side-channel attacks) as shown in Fig.~\ref{progression_of_CS}(a). However, this analog solution faces scalability challenges when transitioning to newer technology nodes. Adaptation to each technology node requires significant engineering effort.
To address this, a synthesis-friendly current source was proposed by Ghosh et al.~\cite{ghosh2021syn}, as depicted in Fig. ~\ref{progression_of_CS}(b). This digital-friendly approach brings the benefits of signature attenuation in the digital domain, maintaining scalability. 
The work by Ghosh et al. ~\cite{ghosh2021syn} employs a PMOS-based power-gate approach for current source utilization. Specifically, a stacked PMOS structure is biased using a self-connected NOT gate, internally generating a voltage of $\frac{V_{DD}}{2}$. This solution effectively addresses the scalability challenge associated with signature attenuation-based countermeasures. However, This architecture uses source-degenerated current source (CS) structures. Source degenerated CS exhibits lower output impedance than the cascoded structure, reducing attenuation.
To overcome this limitation, we propose a digital cascoded current source (DCCS). The DCCS configuration consists of two PMOS transistors, each independently biased, as illustrated in Fig. ~\ref{progression_of_CS}(c). The NOT gate’s output is connected to its input, stabilizing it at $\frac{V_{DD}}{2}$ to bias the lower PMOS. The upper PMOS, on the other hand, is biased using a stack of NAND gates. Specifically, three stages of 16 self-connected NAND gates serve as a resistive divider. By connecting one input of the NAND gate to its output, we incorporate a self-biased structure. Importantly, the NAND gate provides control over the NOT gate. When the other input is ‘1’, it functions as a self-biased inverter, effectively acting as a resistor in the implemented architecture. Conversely, if the other input is ‘0’ (resulting in a NAND output '1'), the NMOS series path is closed, exhibiting high resistance (NMOS in the cut-off region). This controllability via the second input port enables a tunable resistive-divider structure, facilitating the biasing of the upper PMOS.
  \begin{figure}[!t]
  \centering
   \includegraphics[width=0.5\textwidth]{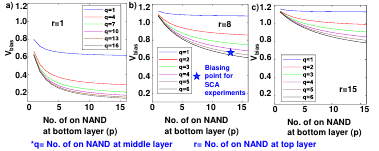}
   \caption{{\color{black} Created biasing voltage by NAND structure when the number of on NAND gate at the top r is a) 1, b) 8, and c)15, respectively. We create variable voltage by biasing the top PMOS of the CS slices using this structure.}}
   \label{vbias_vs_p}
 \end{figure}
We use these two techniques to bias the PMOS transistors, resulting in a synthesizable cascoded current source. {\color{black} Biasing voltage of the top PMOS ($V_{bias}$) is given by following equation. }
{\color{black}
\begin{equation}
\begin{array}{l}
V_{bias}= V_{DD} \times \frac{Z_{\text {bottom }}+Z_{\text {mid }}}{Z_{\text {bottom }}+Z_{\text {mid }}+Z_{to p}}\\[8pt]
=V_{D D} \times \frac{\left(\frac{r_{\text {on }}}{p} \| \frac{r_{o f f}}{16-p}\right)+\left(\frac{r_{\text {on }}}{q} \| \frac{r_{o f f}}{16-q}\right)}{\left(\frac{r_{\text {on }}}{p} \| \frac{r_{\text {off }}}{16-p}\right)+\left(\frac{r_{\text {on }}}{q} \| \frac{r_{o f f}}{16-q}\right)+\left(\frac{r_{\text {on }}}{r} \| \frac{r_{o f f}}{16-r}\right)}
\end{array}
\end{equation}
}
{\color{black} where $Z_{top}$, $Z_{mid}$, and $Z_{bottom}$ are the impedances of different NAND stages, $r_{on}$ \& $r_{off}$ are self-connected and off resistance of a single NAND gate and $p,q,r$ are number of self-connected NAND at bottom, middle, and top stage respectively. We can control the resistance by controlling $p,q,r$. Note that, assuming $r_{off}>>r_{on}$, this structure, ideally, can generate voltages between 0 and $V_{DD}$. However, the contribution of $r_{off}$ restricts the full swing. For example, with 16 stages of minimum-sized NAND gates, we can generate voltage ranging from 110mV to 1.15V when $V_{DD}=1.2V$ as shown in Fig~\ref{vbias_vs_p}(a-c) by using $p,q,r$ as tuning knob. We vary the number of self-connected NAND at every stage of the NAND structure and plot created biasing voltage with different numbers of top self-connected NAND gates($r$). For this work, we use $V_{bias} = 0.72V.$}
  \begin{figure}[!ht]
  \centering
   \includegraphics[width=0.45\textwidth]{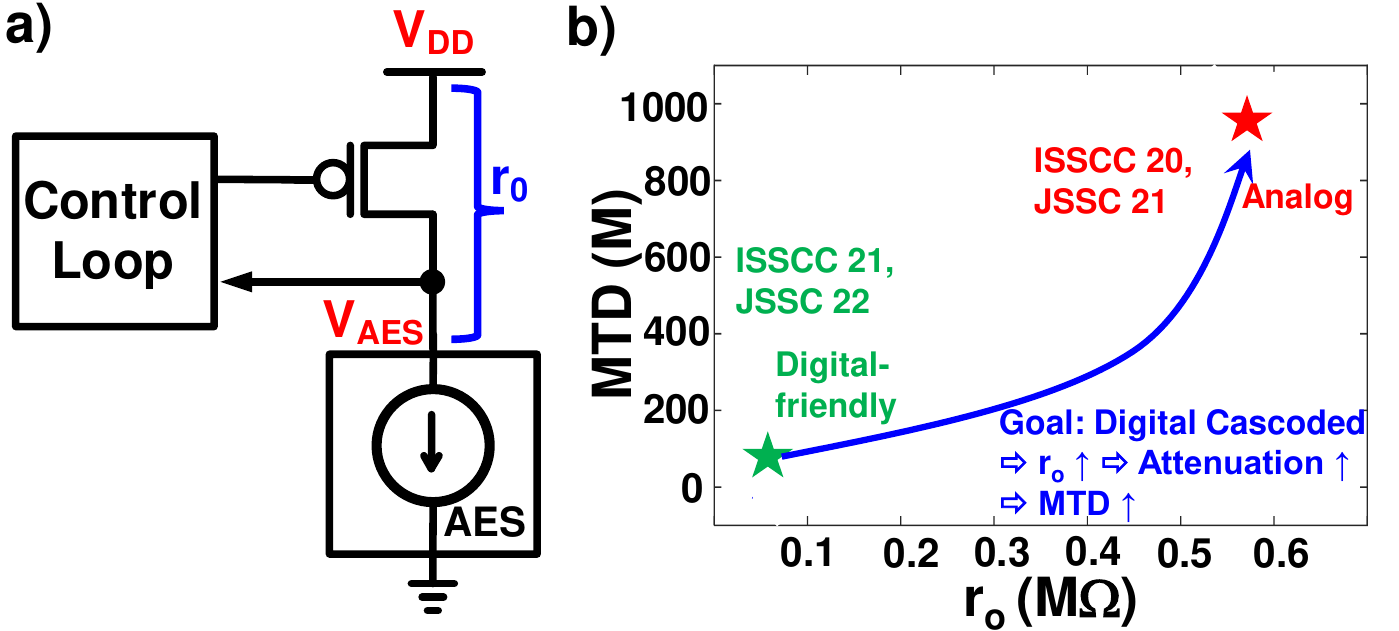}
   \caption{{\color{black} a) LDO architecture for security. b) Higher output impedance ($r_o$) helps achieving higher signature attenuation. This work achieves analog-like $r_o$ by using the digital circuit.}}
   \label{shunt_LDO}
 \end{figure}
This approach maintains scalability while providing substantial attenuation by creating cascoded structure, positioning it as a key component in signature attenuation-based countermeasures. Notably, through parametric extracted simulations, we achieve an impressive 343$\times$ attenuation, surpassing the results reported in Ghosh et al.'s previous work (which achieved 30$\times$ attenuation~\cite{ghosh2021syn}) as shown in Fig.~\ref{progression_of_CS}(d).
{\color{black} This architecture is a LDO architecture (Fig.~\ref{shunt_LDO}(a)) which is proven effective against side-channel attack. Note that the control loop at shunt LDO also uses a shunt path for stable internal voltage, hence providing higher security. Digital cascoded architecture helps us achieve similar high output impedance of analog structure (Fig.~\ref{shunt_LDO}(b)). We will explore MTD improvement through silicon experimentation, which is described in section V.}

  \begin{figure}[!t]
  \centering
   \includegraphics[width=0.48\textwidth]{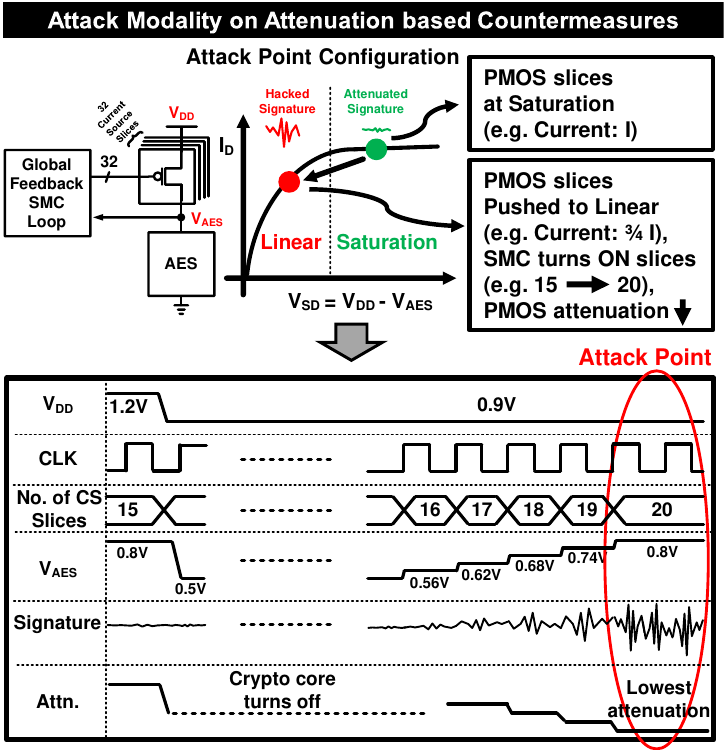}
   \caption{{Attack modality on attenuation-based countermeasure. Manipulating the current source and operating them in the linear region while supplying enough average current for the crypto engine leads to information leakage.}}
   \label{attack_modality}
 \end{figure}
\subsection{Switched Mode Controller (SMC) as Global Feedback Loop  \& Ring Oscillator as Local Negative Feedback}
Our design uses a digital switched-mode controller (SMC) loop as global negative feedback. The adoption of SMC is prevalent in signature attenuation-based solutions, as discussed in ~\cite{ghosh2021syn}. However, an in-depth understanding of this component is crucial for assessing the attack surface against such countermeasures.

The ring oscillator (RO) converts $V_{AES}$ voltage into frequency. RO output undergoes frequency division before being counted by an asynchronous counter. This frequency division ensures low-power operation without sacrificing precision in the asynchronous counter. A decision circuit is also employed to selectively activate or deactivate the current source (CS) slices. This dynamic adjustment responds to variations in average current drawn by the cryptographic engine due to process, voltage, and temperature (PVT) fluctuations or changes in operating frequency.
While the RO is a local negative feedback (LNFB) path, it is not utilized for random noise injection, as demonstrated in Ghosh et al.'s work~\cite{ghosh2021syn}. Our evaluation focuses purely on the signature attenuation technique {\color{black} for a fair comparison of the key technique}. The RO also plays a role in detecting malicious voltage drop-based attacks. For the sake of continuity, we will elaborate on it in section IV.

\section{\color{black} Voltage-drop Linear-region Biasing Attack}
Recent advancements in physical countermeasures introduce novel attack vectors that need exploration. As part of the sustainable evolution of physical security measures within the industry, we delve into the attack landscape, specifically targeting signature attenuation-based countermeasures. Our investigation marks the first exploration of this attack modality in the solid-state circuit community (SSCS). In subsequent subsections, we propose an attack detector to detect such an attack. 
\subsection{Possibility of attack by manipulating global negative feedback loop}
The attack modality is elucidated in Fig.~\ref{attack_modality}, involving manipulating the SMC loop. Consider an encryption engine that draws a current of $15I$, which is supplied by 15 current source (CS) slices operating in the saturation region. Now, through trial and error, an attacker can deliberately reduce the supply voltage ($V_{DD}$) slightly. Due to this abrupt voltage drop, the encryption engine may initially fail to operate. Still, the global negative feedback loop (SMC) will engage, aiding the circuit into a steady state.
  \begin{figure}[!t]
  \centering
   \includegraphics[width=0.5\textwidth]{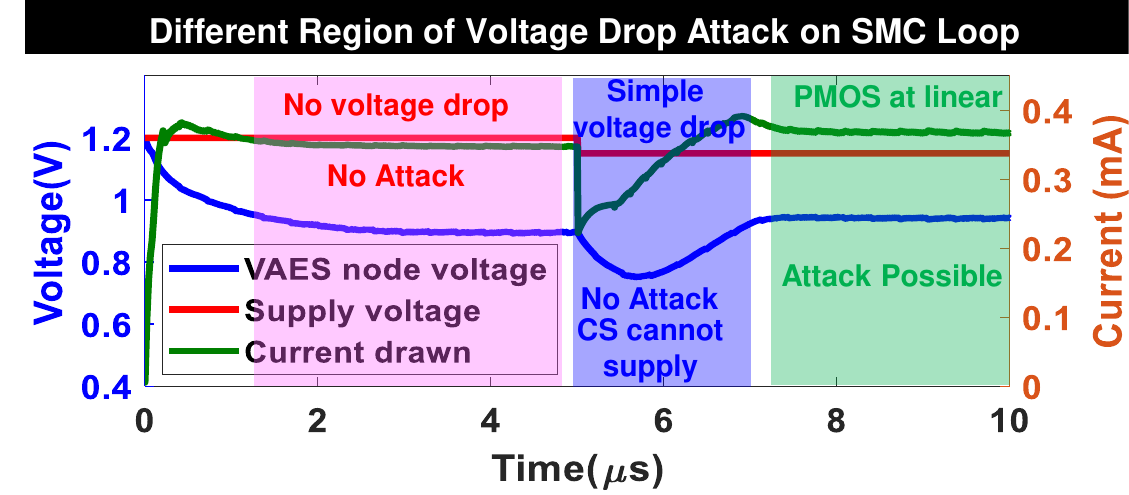}
   \caption{{Different region for voltage drop based attack. At stable $V_{DD}$, there is no voltage drop leading to no attack. Initial voltage drop can not support this attack as CS cannot supply the AES. However, GNFB stabilizes CS slices in linear regions; it will start leaking information. }}
   \label{attack_region}
 \end{figure}
    \begin{figure}[!t]
  \centering
   \includegraphics[width=0.48\textwidth]{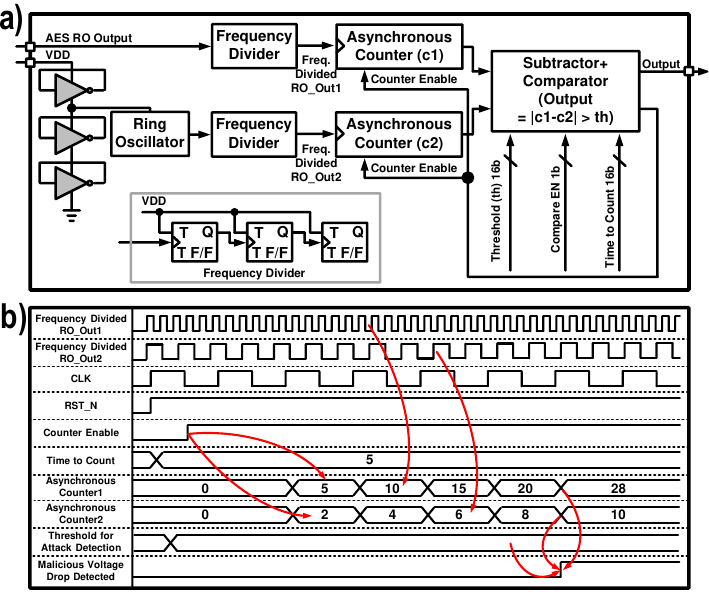}
   \caption{{a) Attack detector circuit for malicious VLB attack. b) Sample waveform of attack detector. }}
   \label{attack_detector}
 \end{figure}
To compensate for the reduced average current, the SMC loop activates additional CS slices. For instance, if each CS slice can provide $\frac{3}{4} \times I$ current, then 20 CS slices would collectively deliver the required $15I$ current for the encryption engine as shown in Fig.~\ref{attack_modality}. Notably, all these slices operate in the linear region, resulting in significantly lower output impedance. Unfortunately, this reduced attenuation leads to heightened information leakage.
The simulated impact of a voltage drop-based attack on the global negative feedback loop is depicted in Fig.~\ref{attack_region}. In the absence of any voltage drop, when the current source (CS) slices operate in the saturation region, there is no vulnerability to attack. The system remains stable, as shown in the red region. As the voltage ($V_{DD}$ node) experiences a slight drop, the SMC loop becomes destabilized. Notably, a significant droop occurs at the $V_{AES}$ node (indicated by the blue region in Fig.~\ref{attack_region}). However, an attack is not feasible in this scenario because the CS cannot supply the required current to the AES.
Consequently, the AES remains non-operational while the SMC becomes active. Eventually, the loop settles back into stability (green region). The CS slices operate in the linear region at this point, creating a lack of high attenuation. Unfortunately, this lack of high attenuation introduces the possibility of an attack.
{\color{black} It is important to note that attack on this countermeasure is not always possible. We need to meet the following circuit criterion to achieve the attack point.}
{\color{black}
\begin{equation}
\begin{array}{l}I_{Crypto} = I_{s a t} \times m=I_{\text {lin}} \times n \\ \Rightarrow k\left(V_{G S}-V_T\right)^2 \times m=\frac{k}{2}\left(V_{GS}-V_T-V_{DS}/2\right)V_{D S}  \times n \\ \Rightarrow \frac{2\left(V_{G S}-V_T\right)^2}{\left(V_{G S}-V_T-V_{D S}/2\right)^2 \times V_{D S}} \times \frac{I_{\text {Crypo }}}{I_{\text {sat }}}=n\\[4pt] \Rightarrow n=\frac{2 I_{\text {crypto }}}{K} \times \frac{1}{\left(V_{G S}-V_T-V_{DS/ 2} )/ V_{D S}\right.} \\\ \ \ \ \ \ =\frac{2}{K} \cdot \frac{I_{\text {Crypto }}}{\left(V_{G S}-V_{T}-\frac{V_{D S}}{2}\right) \times V_{D S}} \leqslant n_{\text {max }} \\\end{array}    
\end{equation}
where $m,n$ are PMOS turned on to supply the crypto engine in saturation and linear region respectively; $K$ MOS device constant, $I_{crypto}$ is average crypto current. $V_{GS}, V_{DS},$ and $V_T$ are absolute gate-to-source, drain-to-source and threshold voltage respectively. $I_{sat}$ and $I_{lin}$ are saturation and linear region current of single PMOS gates. $n_{max}$ is maximum current source slices. Note that if all the current sources combined cannot drive the crypto core, the attack will not be successful. 
}
  \begin{figure}[!ht]
  \centering
   \includegraphics[width=0.5\textwidth]{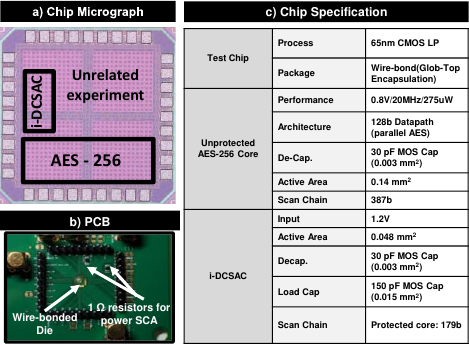}
   \caption{{a) IC micrograph b) PCB for testing c) IC specification.}}
   \label{chip_micrograph}
 \end{figure}
   \begin{figure}[!t]
  \centering
   \includegraphics[width=0.5\textwidth]{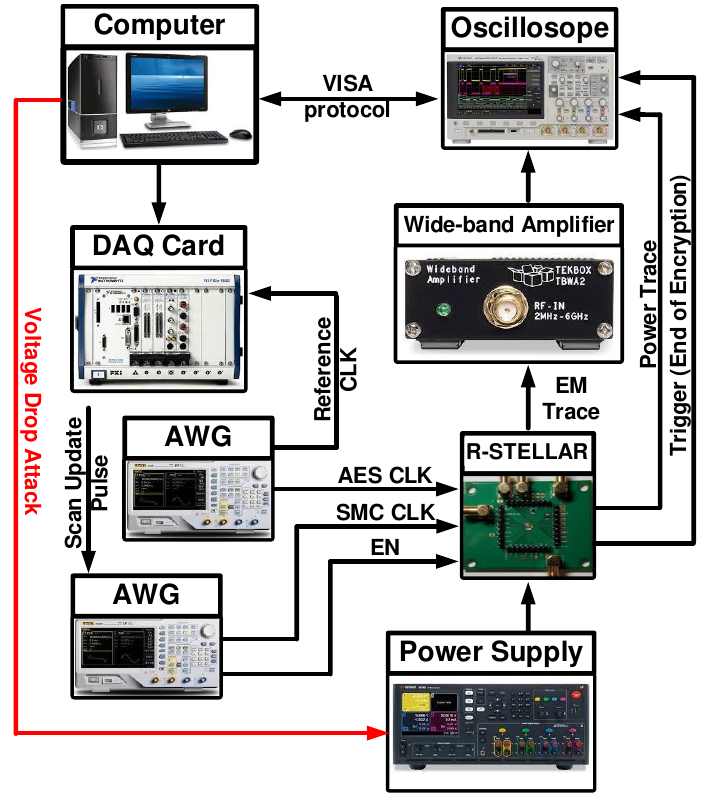}
   \caption{{Measurement setup for power/EM side channel.}}
   \label{attack_setup}
 \end{figure}
     \begin{figure}[!t]
  \centering
   \includegraphics[width=0.5\textwidth]{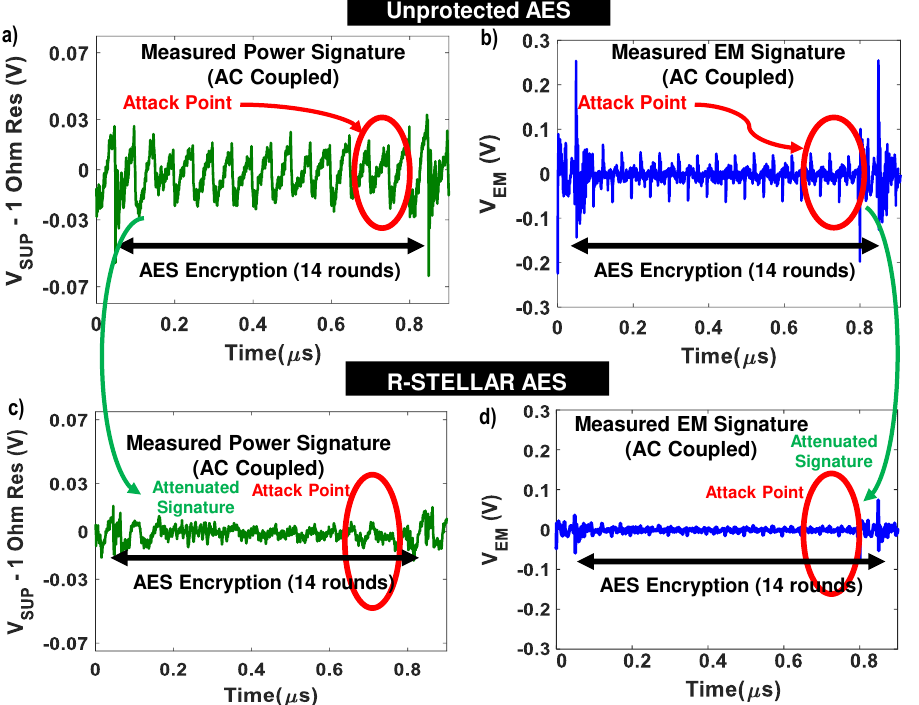}
   \caption{{Time domain trace for different configurations: a) unprotected power, b) unprotected EM, c) protected power, d) protected EM. }}
   \label{time_domain_trace}
 \end{figure}
    \begin{figure*}[!h]
  \centering
   \includegraphics[width=0.98\textwidth]{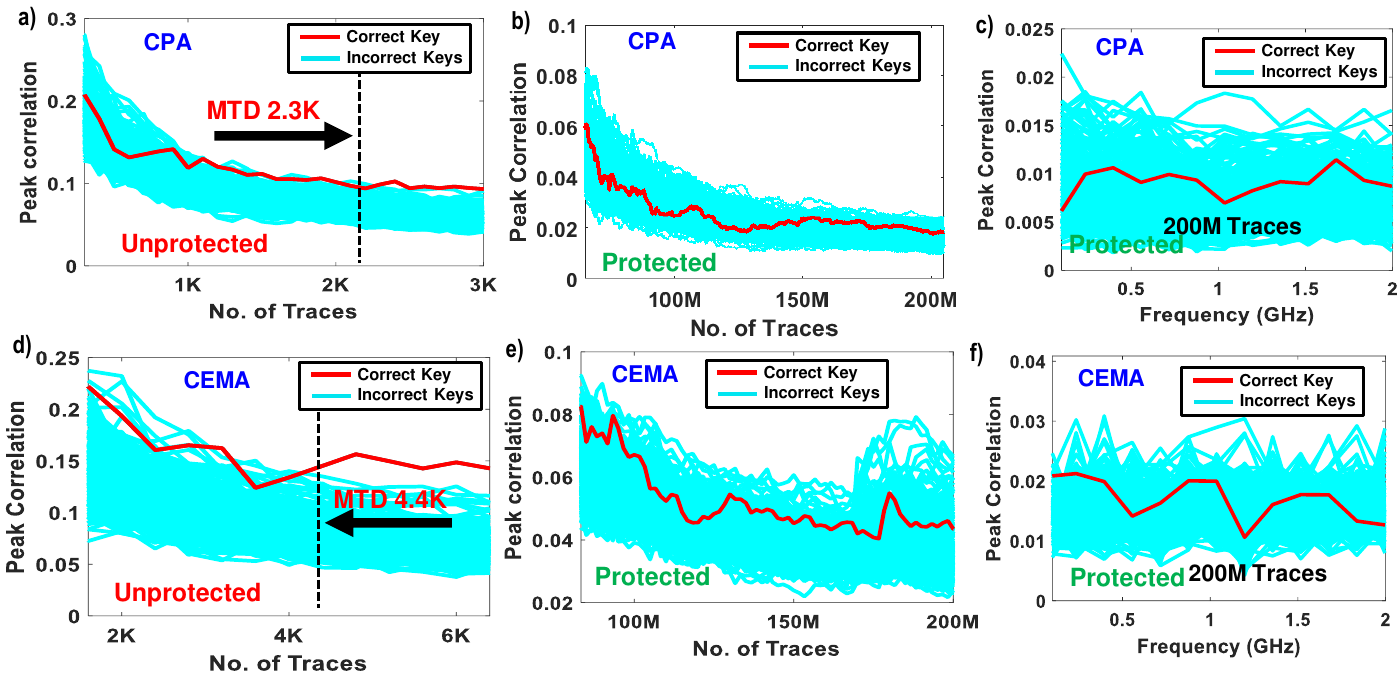}
   \caption{{Correlational power attack (CPA) on a) unprotected AES-256 b) protected AES-256 c) frequency domain CPA on protected AES-256. Correlation EM attack (CEMA) on d) unprotected AES-256 e) protected AES-256 f) frequency domain CEMA on protected AES-256.)}}
   \label{CPA_CEMA}
 \end{figure*}

\subsection{VLB Attack Detector}
We introduce an attack detection circuit to mitigate malicious VLB attacks on signature attenuation countermeasures. The circuit, depicted in Fig.~\ref{attack_detector}(a), aims to identify the voltage discrepancy between $V_{DD}$ and $V_{AES}$, enabling successful detection of malicious attacks.\\
Within our system, the LNFB employs an RO to stabilize the $V_{AES}$ node, serving as an input to the GNFB. We utilize the same RO as a critical component to ensure the sustainability of our signature attenuation-based countermeasure. The RO output undergoes frequency division and feeds into an asynchronous counter, yielding an estimation of the AES voltage. 
Additionally, we employ another ring oscillator to estimate the global $V_{DD}$. By dividing the voltage using stacked inverters, we achieve approximately $\frac{2}{3}$ of the global $V_{DD}$. This voltage division strategy ensures that both the counted numbers remain closely aligned. The divided voltage is digitized through a replica frequency divider and an asynchronous counter. Subsequently, both counted values are input to a digital comparator, which functions as the voltage drop detector.
Ideally, the difference between these two numbers should be minimal, given the similarity between the voltage-divided $V_{DD}$ and $V_{AES}$. This comparison is configurable, allowing us to adjust the estimated difference using scan chain within the voltage drop detector circuit. In the event of a voltage drop-based attack, where $V_{DD}$ is intentionally reduced, the difference between the counter outputs surpasses a predefined threshold. This occurrence signals the possibility of VLB side-channel attack on our signature attenuation-based countermeasures, ultimately activating protective measures, including halting the encryption engine. \\
Fig. ~\ref{attack_detector}(b) illustrates the working principle of the attack detector. Frequency divided RO outputs (RO\_Out1 deduced from $V_{DD}$ and RO\_Out2 deduced from $V_{AES}$) are counted using an asynchronous counter when the counter enable signal is high. 'Time to count' determines the time required (\# clock cycles) to accumulate RO outputs before calculating the difference between them. In this example, asynchronous counter1 and asynchronous counter2 accumulate the RO output for five clock cycles, which are 20 \& 8, respectively. The difference of 12 is greater than the expected threshold of 10, which indicates the ongoing malicious voltage drop attack. 
\\
\section{Measurement Results}\label{sec:measurement_results}
In this section, we delve into the measurement results. Initially, we outline the IC specifications, followed by a detailed discussion of the measurement setup. Finally, we explore various security evaluations related to the implemented countermeasure.
\subsection{IC specification}
The integrated circuit (IC) micrograph is depicted in Figure ~\ref{chip_micrograph}(a). This $1 mm^2$ IC features an AES-256 crypto-engine as a use case. Notably, the left side of the IC houses the implemented countermeasure, R-STELLAR. A sample PCB is also shown in Figure ~\ref{chip_micrograph}(b). 1 $\Omega$ resistor is used in $V_{DD}$ series path to sense the current for power side-channel attack.
The IC specifications are summarized in Fig.~\ref{chip_micrograph}(c). Fabricated using the TSMC 65nm CMOS LP process, the IC employs chip-on-board packaging with glob-top encapsulation. Load characterization is performed on the unprotected core, as illustrated in Figure ~\ref{full_system_archi}(c). The parallel 128-bit datapath AES serves as the crypto engine (Fig. ~\ref{full_system_archi}(b)), operating at 20MHz and 0.8V $V_{DD}$. At this configuration, AES-256 consumes 275.2uW of power. A 30pF decoupling capacitor (moscap) is placed, occupying an area of 0.003$mm^2$. The total active area of the encryption engine is 0.14$mm^2$.
 \begin{figure}[!ht]
  \centering
   \includegraphics[width=0.3\textwidth]{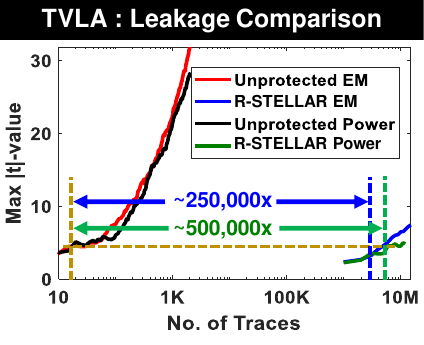}
   \caption{{TVLA-based leakage analysis for all configurations. }}
   \label{tvla}
 \end{figure}
The countermeasure occupies an active area of 0.048$mm^2$. To further stabilize the $V_{AES}$ node and provide resilience against large droops, an additional load capacitor of 150pF is incorporated. This capacitor, occupying an area of $0.015mm^2$, contributes to area overhead significantly. R-STELLAR operates with a 1.2V $V_{DD}$ input. Protection needs 179 bits of scan chain for configuration, although some of these scan bits are also utilized for unrelated experiments within the same die.
 \begin{figure}[!t]
  \centering
   \includegraphics[width=0.5\textwidth]{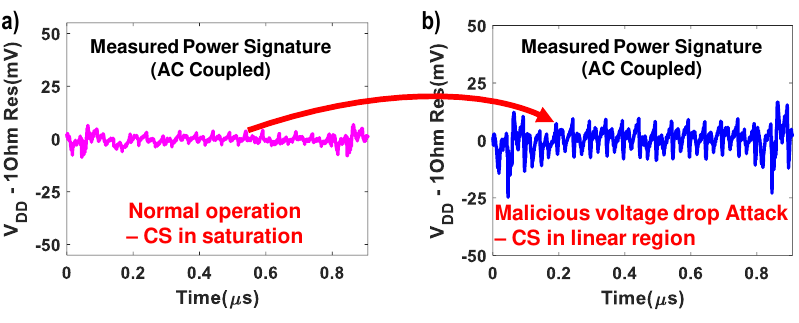}
   \caption{{AC coupled power trace is observed for a) protected AES-256 b) protected AES-256 under malicious voltage drop based attack. }}
   \label{time_domain_trace_for_attack}
 \end{figure}
    \begin{figure}[!h]
  \centering
   \includegraphics[width=0.48\textwidth]{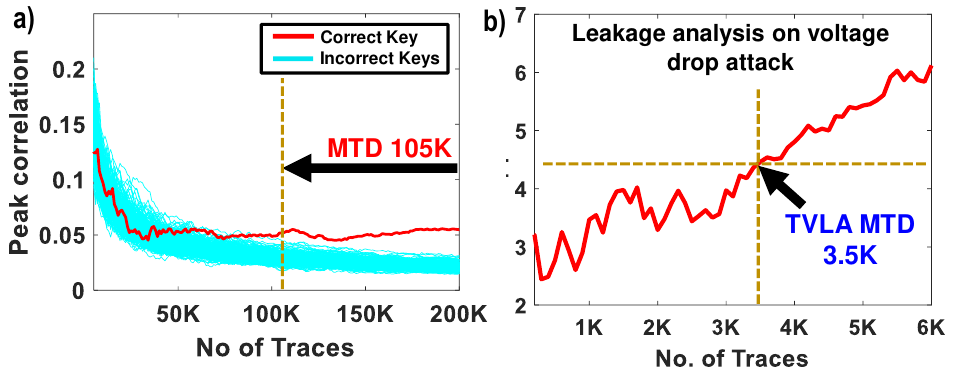}
   \caption{{a) MTD is reduced to 105K by malicious vVLB attack. b) TVLA MTD (traces required to reach max $|t|-value$ of 4.5) reduces to 3.5K. }}
   \label{mtd_after_attack}
 \end{figure}
\subsection{Measurement setup}
The attack setup is depicted in Fig.~\ref{attack_setup}. A power trace is acquired using a 5 GSps oscilloscope, while an H-probe with a 10 mm diameter is employed for electromagnetic (EM) trace collection. The EM trace is subsequently amplified using a wideband amplifier before being acquired through the oscilloscope. The end of encryption is indicated by a trigger signal, aiding in the alignment of the collected traces. These traces are then transmitted to a computer via the VISA protocol for further processing. The computer utilizes a NI-Data Acquisition (NI-DAQ) card to configure the integrated circuit (IC).
Additionally, an arbitrary waveform generator (AWG) supplies the IC with enable, reset, and clock signals. Typically, a stable power supply powers up the IC. But here, we control the supply from the computer to introduce VLB attack.

\subsection{Correlational Power/EM Attack \& Leakage Analysis}
The time domain measurement results are depicted in Fig.~\ref{time_domain_trace}. In Fig.~\ref{time_domain_trace}(a), the AC-coupled power trace is displayed, showing 14 cycles of AES operation. Fig.~\ref{time_domain_trace}(b) presents the amplified AC-coupled electromagnetic (EM) trace. Additionally, Fig.~\ref{time_domain_trace}(c) shows the attenuated power trace. 
Finally, Fig.~\ref{time_domain_trace}(d) displays the attenuated EM traces. Attenuation is clearly visible when CS is operating in the saturation region. 

We have chosen the Hamming Distance between the last two rounds as our attack model. The correct key is revealed within 2.3K traces in the standard correlational power attack, as depicted in Fig.~\ref{CPA_CEMA}(a) for the unprotected implementation. However, in the presence of R-STELLAR, the correct key is not revealed even after analyzing 200M traces (Fig.~\ref{CPA_CEMA}(b)). To further validate our findings, we conducted a frequency domain correlation power analysis (CPA) over a frequency range of 100 MHz to 2 GHz (Fig.~\ref{CPA_CEMA}(c)). No peak correlation is detected across the entire spectrum.
It’s worth noting that attackers often attempt to mitigate the effects of noise by averaging traces, thereby increasing the signal-to-noise ratio (SNR). Our attack setup follows a similar approach, employing an averaging factor of 1000 during the attack. In contrast, standard CEMA (Correlation Electromagnetic Analysis) with the Hamming Distance between the last two rounds successfully reveals the correct key using just 4.4K traces (Fig.~\ref{CPA_CEMA}(d)). No correct key byte is exposed even after analyzing 200M traces using CEMA (Fig.~\ref{CPA_CEMA}(e)).
We performed frequency domain CEMA on a protected AES implementation to ensure security in the frequency domain. No key byte is revealed when measured with 200M traces across the entire frequency spectrum of 100 MHz to 2 GHz (Fig.~\ref{CPA_CEMA}(f)). \\
TVLA-based leakage analysis was conducted on both unprotected and protected implementations. The $|t|-value$ was calculated using fixed and random plaintexts. A $|t|-value$ exceeding 4.5 indicates the presence of a leaky component. The unprotected implementation starts to leak within 100 traces for both the power and EM side channels. The countermeasure, R-STELLAR, shows the presence of leakage after 2.5M and 5M traces for EM and power SCA, respectively, as shown in Fig.~\ref{tvla}. This is $~250,000\times$ and $500,000\times$ improvement with respect to unprotected implementation.   
 \begin{figure}[!t]
  \centering
   \includegraphics[width=0.45\textwidth]{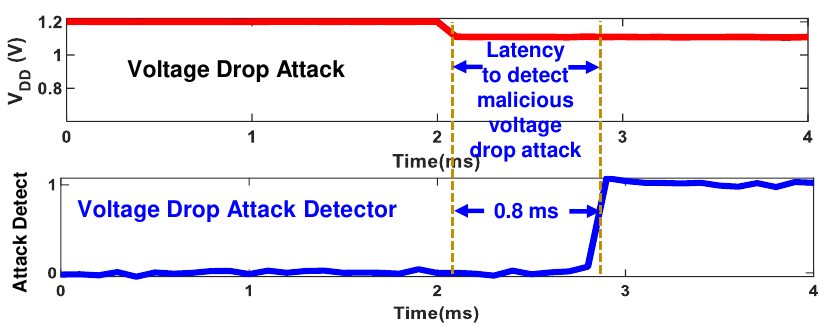}
   \caption{{Attack detector can detect such attack within 0.8ms.}}
   \label{attack_detection_time_domain}
 \end{figure}
    \begin{table*}[!ht]
  \centering
   \includegraphics[width=0.9\textwidth]{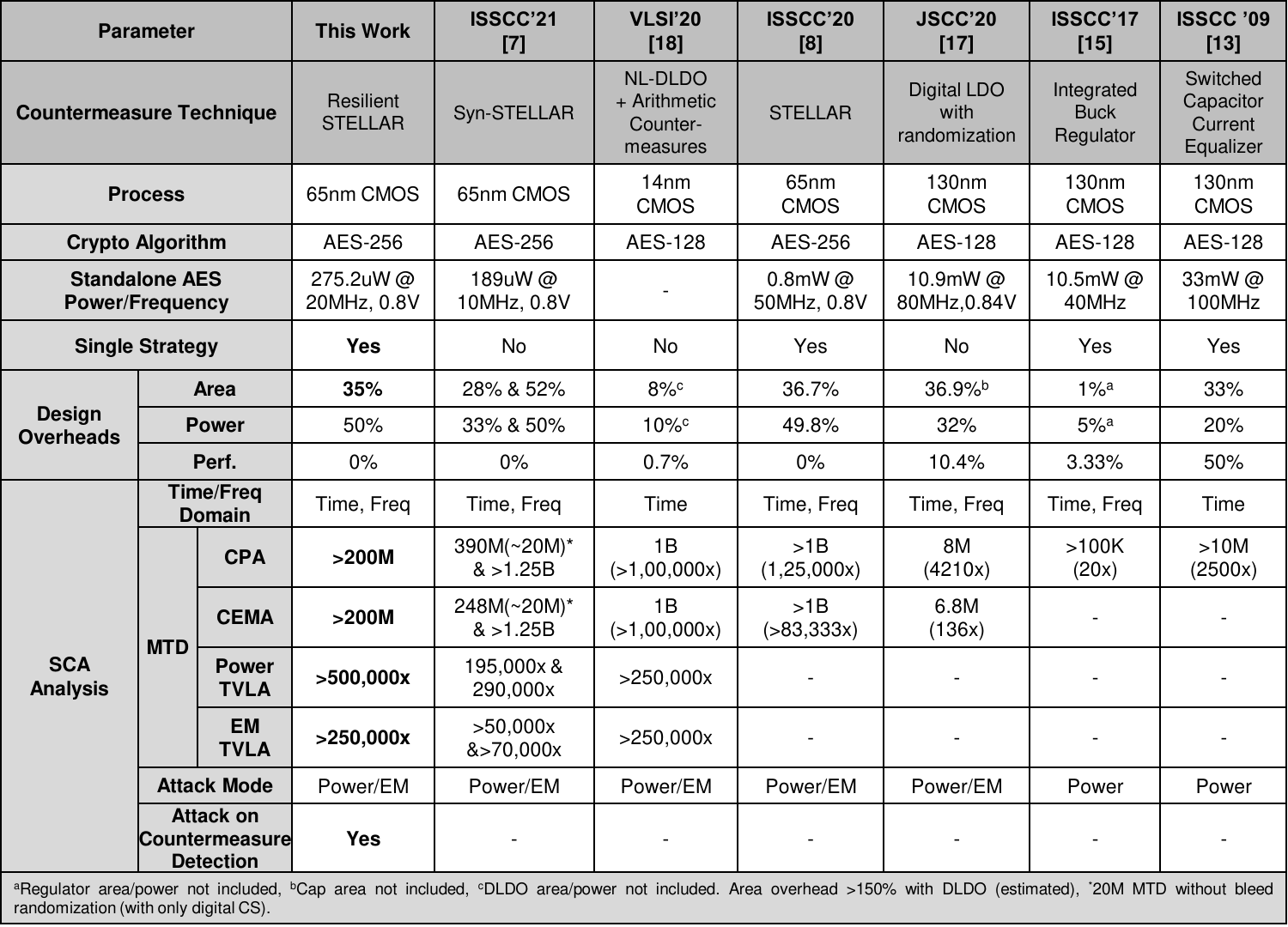}
   \caption{{Comparison with respect to other state-of-the-art.}}
   \label{comparison_table}
 \end{table*}
\subsection{Malicious Voltage Drop-based Attack \& Mitigation}
For the first time, we explore a dedicated side-channel attack on a physical countermeasure. Our approach leverages a malicious voltage drop-based attack, which reduces the attenuation that the implemented power delivery circuit provides. Specifically, the PMOS begins to operate in the linear region due to a slight voltage drop at $V_{DD}$ node.
Fig.~\ref{time_domain_trace_for_attack} presents the measured time domain trace. Notably, the amplitude of the power trace significantly increases compared to steady-state operation (Fig.~\ref{time_domain_trace_for_attack}(b) vs. Fig.~\ref{time_domain_trace_for_attack}(a)). Following the malicious voltage drop, we incorporate a CPA. The correct key byte is retrieved with just 105K traces (Fig.~\ref{mtd_after_attack}(a)).
We perform a frequency domain CPA using 150K traces to validate our findings further. The results confirm the presence of leaky components at 400 MHz. Additionally, we conduct a TVLA (Test Vector Leakage Assessment) based leakage analysis, revealing that meaningful information leakage begins from just 3K traces (Fig.~\ref{mtd_after_attack}(b)).
The proposed mitigation technique effectively detects the described attack within a time frame of 0.8 ms, achieving 100\% accuracy, as illustrated in Fig.~\ref{attack_detection_time_domain}. Notably, the short detection time ensures the countermeasure's robustness. 
{\color{black} Assuming reduced MTD of $105K$, AES operating at 20MHz, and 14 cycles of operations, a SCA attack can be successful within 73.5ms assuming 0 oscilloscope capture time. Notably, 0 oscilloscope capture time is unrealistic. Nevertheless, our approach detects an attack-on-countermeasure within 0.8ms. Only 1.1\% of the encryptions are possible in this time frame, eliminating the possibility of attack.} 
The latency for attack detection is also influenced by the clock period of the attack detector. While we typically operate at a low frequency (10 KHz) due to the slow nature of the SMC loop, a faster clock could further enhance the detection speed if needed in future scenarios. Table~\ref{comparison_table} provides a comparative analysis between our proposed work and existing state-of-the-art techniques. 
\section{Conclusion}\label{conclude}
Our approach offers a scalable physical countermeasure while maintaining high security as a standalone technique. Importantly, no attacks have been explored on these countermeasures to date. In this work, we investigate an attack on the countermeasure circuit for the first time and introduce a detector circuit to identify such attacks.
In summary, our work achieves over 200M MTD with synthesizable signature attenuation as a single countermeasure technique. Additionally, we explore an attack modality in the presence of physical countermeasures, specifically focusing on synthesizable signature attenuation. Our proposed method effectively detects supply voltage drop-based linear-region biasing attacks within less than 1 ms. Practical CPA within this time range is infeasible. Furthermore, this generic countermeasure can be cascaded with other algorithmic or architectural countermeasures to enhance overall security.

\bibliographystyle{unsrt} {
    \bibliography{JSSC.bib}
}

\end{document}